\documentclass[aps,prx,twocolumn]{revtex4-1}
\usepackage{amsmath}
\usepackage{amsfonts}
\usepackage{graphicx}
\usepackage{makecell}
\usepackage{xcolor}

\usepackage{bm}

\begin{document}

\title{Magnetic moment generation in small gold nanoparticles via the plasmonic inverse Faraday effect}
\author{J\'{e}r\^{o}me Hurst}
\author{Peter M.\ Oppeneer}
\affiliation{Department of Physics and Astronomy, Uppsala University, P.\ O.\ Box 516, SE-75120 Uppsala, Sweden}
\author{Giovanni Manfredi}
\author{Paul-Antoine Hervieux}
\affiliation{ Universit\'{e} de Strasbourg, CNRS, Institut de Physique et Chimie des Mat\'{e}riaux de Strasbourg, UMR 7504, F-67000 Strasbourg, France}
\date{\today}

\begin{abstract}
We theoretically investigate the creation of a magnetic moment in gold nanoparticles by circularly polarized laser light. To this end, we describe
the collective electron dynamics in gold nanoparticles 
using a semiclassical approach based on a quantum hydrodynamic model that incorporates the principal quantum many-body and {nonlocal} effects, such as the {electron spill-out}, the Hartree potential, and the exchange and correlation effects. {We use}
a variational approach 
to investigate the breathing and the dipole dynamics induced by an external electric field.
We show that gold nanoparticles can build up a static magnetic moment {through the} interaction with a circularly polarized laser field.
{We analyze that} the {responsible}
physical mechanism is a {plasmonic, orbital} inverse Faraday effect, {which} can be understood {from}
the time-averaged electron current that contains currents rotating on the nanoparticle's surface. The computed laser-induced magnetic moments are sizeable, of about  0.35 $\mu_{\rm B}$/atom for a laser intensity of $45 \times 10^{10}$ W/cm$^2$ at plasmon resonance.

\end{abstract}



\maketitle

\section{Introduction}\label{sec:intro}

The field of magnetoplasmonics has stimulated a {large} amount of scientific interest over the past few {decades} both for reason of fundamental curiosity and in view  of potential technological applications \cite{Armelles2013,Bossini2016,Maksymov2016}. The {basic} principle of this new field of research is to use plasmonic properties to enhance and tune the magneto-optical response. For instance,  it has been shown \cite{Jain2009} that the Faraday rotation in gold-coated maghemite nanoparticles can be enhanced {owing} to the plasmonic properties of gold. {Another example is the observation of a tunable magneto-optical response from nickel nano-disks that can be adjusted by the phase of  localized plasmons \cite{Vavassori2013}.}
Conversely, {magnetoplasmonics can also be used to modulate} the plasmonic properties of metals with an external magnetic field. This has been shown for instance in {Ref.}\ \cite{Melander2012}, where the transmission {of} light through a thin metal film with a periodic sub-wavelength hole array {could} be manipulated via an externally applied magnetic field. {Also, magnetic-field induced modulation of circular magnetoplasmonic modes has been demonstrated for gold nanoparticles by means of magnetic circular dichroism spectroscopy \cite{Pineider2013}.}

The plasmonic properties are mediated by plasmons and result from the coupling between an electromagnetic wave and a collective oscillation of the surface free charges at the interface between two media with permittivities of opposite signs, typically a dielectric and a metal. A well-known example of plasmons are the localized surface plasmons \cite{Petryayeva2011} in {a} gold nanoparticle, {which} correspond to oscillations of the electron cloud along the direction of the applied electric field. This leads to a strong enhancement of the electric field at the surface of the nanoparticle due to the charge displacements.

Plasmonic systems are of great interests since they strongly interact with light. For instance they can be used to focus the light in a small region of space leading to a strong local enhancement of the laser field  \cite{Kreibig1995}. 
Moreover, it is well known that light carrying spin angular momentum can couple it into the electronic system through the inverse Faraday effect \cite{Vanderziel1965,Pershan1966}. This is a nonlinear optical effect that is characterized by the creation of a static induced magnetization that is proportional to the laser intensity \cite{Pershan1966}. The inverse Faraday effect  has very recently drawn attention as a possible pathway to enable fast, all-optical switching of the magnetization in a ferri- or ferromagnetic material \cite{Stanciu2007,Alebrand2012,Mangin2014,Lambert2014,Elhadri2016,John2017,Vomir2017}. The magnetization induced by a circularly-polarized laser pulse acts on the equilibrium magnetization and thereby effectuates its switching. The inverse Faraday effect has recently been  investigated for many systems such as metals \cite{Hertel2006,Kurkin2008,Woodford2009,Popova2011,Berritta2016},  molecular magnets \cite{Tokman2009}, and plasmonic systems \cite{Smolyaninov2005,Nadarajah2017,Koshelev2015,Hamidi2015}. In order to make the switching more efficient the induced magnetization has to be as large as possible. It is currently being investigated whether this can be realized by plasmonic antennas or nanoparticles \cite{Hamidi2015,Liu2015,Dutta2017,Kataja2018}. In that sense, plasmon could be used to enhance the conversion of light angular momentum into electronic angular momentum, opening the possibility for ultrafast plasmon-assisted all-optical switching.

In this paper, we focus on the inverse Faraday effect in gold nanoparticles. The latter are known to support strong plasmonic effects \cite{Kreibig1995}. We use a quantum hydrodynamic (QHD) model  \cite{Manfredi2005,Vladimirov2011,Ciraci2016} to describe the interaction between the surface plasmons in gold nanoparticles and a circularly polarized {laser field}. QHD models are orbital-free methods that are used to study the dynamics of large systems including some quantum effects and many-body interactions. Such models were recently used to model the electron dynamics in thin films \cite{Crouseilles2008}, metallic nanostructures \cite{Ciraci2017,Hurst2014,Toscano2015,Hurst2016}, semiconductor quantum wells \cite{Haas2009} and molecular systems \cite{Brewczyk1997}.  Many recent studies have emphasized the importance of spatial nonlocal effects in the optical response of plasmonic systems \cite{David2011,Toscano2015,Raza2015,Ciraci2017,Krasavin2018}. The latter are suitably incorporated in the QHD model through the self-consistent fields and lead to spatial variations of the electron density.

The paper is organized as follows. In {Sec.\ II}, we describe the system and the QHD model. In {Sec.\ III}, we present the results that we {have} obtained for the {laser-}generation of a static magnetization in gold nanoparticles. In {Sec.\ IV}, we propose an explanation of the mechanism responsible for the inverse Faraday effect in gold nanoparticles. In {Sec.\ V}, we study the influence of the nanoparticle size and the laser intensity on the inverse Faraday effect.

\section{Description of the model}

We consider spherical gold nanoparticles with a radius $r_c$ and composed of $N$ ions and $N$ electrons. In our simulations $r_c$ will be on the order of $1-2.5$ nm.
Both parameters are related  by $r_c = r_{s} N^{1/3}$, where $r_{s}$ is the so-called Wigner-Seitz radius. We use the following value for gold $r_s =3.01$ {$a_0$ (Bohr radii)}. We work in the framework of the jellium approximation, i.e.\ we consider that {the} ions are fixed and homogeneously distributed. Thus the ion density is given by $n_i = n_{0} = 3N/(4 \pi r_c^{3})$ inside the cluster and zero outside.
This assumption is justified by the fact that there is a timescale separation between the ion and the electron dynamics. The timescale for the electrons is given by the plasma frequency $\omega_{p} = ({4\pi n_0})^{1/2}$. In the case of gold we obtain a timescale of the order of $1$ femtosecond.
{Here and {henceforth}, all equations will be written in atomic units.}

The electron dynamics is described by the QHD equations \cite{Manfredi2005,Vladimirov2011}, that are derived in a standard way from the kinetic equation (Wigner-Poisson) averaging the electron distribution
function over different velocity moments and choosing appropriate closure relations. Its validity is limited to systems that are large compared to the Thomas-Fermi screening length $\lambda_{\rm F} = v_{ \rm F} /\omega_p$, where $v_{\rm F} = (3\pi^2n_0)^{2/3}$ is the Fermi velocity. The QHD equations reads:
\begin{align}
&\frac{\partial n}{\partial t} + \bm{\nabla}\!\cdot(n\bm{u}) =  0, \label{eq:continuity gold}\\
&\frac{\partial \bm{u}}{\partial t}+\bm{u}\cdot\!\bm{\nabla} \bm{u} = 
 -\bm{E} + \bm{\nabla} V_{\textrm{H}} - \bm{\nabla}\!\,V_{X} - \bm{\nabla}\!\,V_{C} \nonumber \\
& \quad \quad \quad \quad \quad \quad - \frac{\bm{\nabla} P}{n}+\frac{1}{2}\bm{\nabla} \left(\frac{\bm{\nabla}^2\sqrt{n}}{\sqrt{n}}\right), \label{eq:momentum gold}\\
&\bm{\nabla}^2 V_{\textrm{H}}= 4 \pi \left(n - n_{i} \right). \label{eq:poisson gold}
\end{align}
In Eqs.\eqref{eq:continuity gold}-\eqref{eq:poisson gold}, $n(\bm{r},t)$ is the electron density, $\bm{u}(\bm{r},t)$ is the electron mean velocity, and  $V_{\textrm{H}}(\bm{r},t)$ is the Hartree potential. The latter corresponds to the mean-field part of the electron-electron interactions and is a solution of the Poisson equation \eqref{eq:poisson gold} that corresponds to the quasi-static limit of the Maxwell equations. Such an approach neglects retardation effects and is
valid when the size of the nanostructure is much smaller than the light wavelength, which is the case in this study. Equation \eqref{eq:continuity gold} is a continuity equation that represents the conservation of the number of electrons in the system. Equation \eqref{eq:momentum gold} is an Euler equation that provides the evolution of the electron mean velocity under the action of the different forces that appear on the right-hand side. The electric field $\bm{E}$ corresponds to the laser excitation and
the potential $V_{X} {(\bm{r},t)}$ represents the exchange interaction,
\begin{equation}
V_{X} = - \frac{(3\pi ^{2})^{1/3}}{\pi} n^{1/3} -\frac{4\beta}{3}  \frac{\left( \nabla n \right)^{2}}{n^{7/3}} + 2 \beta \frac{\left( \nabla n \right)^{2}}{n^{4/3}},
\end{equation}
where the first term is the local density approximation (LDA) and the other two terms are gradient corrections. The prefactor $\beta$ is a free parameter that we set equal to $\beta = 0.005$, which is a best-fit value frequently used in atomic-structure calculations \cite{Becke1988}. For the correlations, we use the functional proposed by
Brey \textit{et al.} \cite{Brey1990}, which yields the following correlation potential,
\begin{equation}
V_{C} = -\gamma \ln \left[1+\alpha n ^{1/3} \right],
\end{equation}
with $\gamma = 0.03349$ and $\alpha = 18.376$.
The quantity $P$ is a pressure term, {for which} we use the standard expression of the Fermi pressure of a zero-temperature electron gas,
\begin{equation}
P = \frac{1}{5}\left(3\pi ^{2}\right)^{2/3}n^{5/3},
\label{fermi_pressure}
\end{equation}
which is an acceptable approximation {since} the Fermi temperature for metals is much larger than ordinary temperatures (e.g., for gold, $T_{\rm F} = 64\,200$ K).
The last term in Eq.\ \eqref{eq:momentum gold}, often referred to as the von Weizs\"{a}cker correction or the Bohm potential, takes into account quantum diffraction effects. Details about the derivation of the QHD model can be found in Refs.\ \cite{Manfredi2005,Vladimirov2011,Ciraci2016,Michta2015}.

A full resolution of the QHD model is a {complex numerical problem that we do not attempt to solve here. Instead,} we will follow the same method {that was developed earlier} in Ref.\ \cite{Hurst2014}, which is based on a variational approach to the QHD equations. {The full details of the method can be found in the original work}. The QHD model \eqref{eq:continuity gold}-\eqref{eq:poisson gold} can be derived from the following Lagrangian density:
\begin{equation}
\begin{split}
\mathcal{L}_{D}(\bm{r},t)
& \!= \!
n \! \left[ \frac{\partial S}{\partial t} + \frac{\left(\bm{\nabla} S \right)^{2}}{2} \right] \! + \frac{\left(\bm{\nabla} n \right)^{2}}{8 n} \! + \! \frac{3}{10} \left( 3 \pi ^{2} \right)^{2/3} n^{5/3}  \\
& \quad -{}\!\! \frac{3}{4 \pi} \left( 3 \pi ^{2} \right)^{1/3} n^{4/3} -
 \beta \frac{\left(\bm{\nabla} n \right)^{2}}{n^{4/3}}   \\
& \quad - \frac{\left(\bm{\nabla} V_{\textrm{H}} \right)^{2}}{8 \pi} \left(n_{i}   - n \right) V_{\textrm{H}} - n V.
\end{split}
\label{Lagrangian density gold nanoparticles}
\end{equation}
The Lagrangian density depends on three dynamical fields: the electron density $n(\bm{r},t)$, the Hartree potential $V_{\rm H} (\bm{r},t)$, and $S(\bm{r},t)$ which is related to the electron mean velocity as follows: $\bm{u}(\bm{r},t) = \bm{\nabla} S(\bm{r},t)$. The fields $V_{\rm H}(\bm{r},t)$ and $\bm{u}(\bm{r},t)$ are determined by the electron density via the Poisson and the continuity  equation. {The laser field is described in the dipole approximation by the following electric potential: $V = - \bm{r} \cdot \bm{E}$.}

The idea of the variational approach consists in using a particular Ansatz for the electron density in order to compute exactly the Lagrangian density of the system. The Ansatz should reproduce {with good approximation} the correct electron density obtained with \textit{ab initio} techniques. {The chosen Ansatz should contain {a few} time-dependent variables [e.g., the center of mass of the electron cloud $\bm{d}(t)$] in order to describe the dynamics of the system. Next, one computes the Lagrangian $L(t)$ of the system by integrating the Lagrangian density over space, $L(t) \propto \int \mathcal{L}_D(\bm{r},t) d\bm{r}$. Finally,} using the standard Euler-Lagrange equations, one  obtains a set of differential equations for the dynamical variables introduced in the Ansatz of the electron density. Nevertheless, in order to derive a tractable system of equations, one needs to perform the integration of the Lagrangian density in an exact way. This {puts restrictions on the applicability} 
of the method, because the parameterization of the electron density cannot be too complicated.

In Ref.\ \cite{Hurst2014}, the authors found an acceptable Ansatz for the electron density that allows one to perform all the calculations in an exact fashion. They introduced two dynamical variables $d_z(t)$ and $\sigma(t)$ that represent respectively the center of mass of the electrons along the $z$ axis and the spreading of the electron density at the surface of the nanoparticle, an effect known as the spill-out \cite{Kreibig1995}. 
 Since the ions are frozen, a motion of the center of mass of the electrons leads to the creation of an electric dipole. In contrast, the time evolution of $\sigma(t)$ corresponds to an isotropic extension or compression of the electron gas, which is also known as a breathing motion.
{Under these assumptions,} the authors were able to describe the dipole and the breathing dynamics of the electron gas for a laser excitation that was linearly polarized along the $z$ direction. If we now want to consider a laser excitation that is circularly polarized, then one needs to introduce an additional time-dependent variable, {namely} the center of mass along the $y$ direction $d_{y}(t)$. In this work we always consider a laser field that propagates in the $x$ direction with an electric field that is polarized in the $y-z$ plane.
By generalizing the formula found in Ref.\ \citep{Hurst2014}, the new Ansatz for the electron density reads:
\begin{align}
n\left(\bm{r},t\right) =\frac{A}{1 +\exp\left[\left(\frac{s\left(\bm{r},t\right)}{\sigma \left(t\right)}\right)^{3} - \left(\frac{r_c}{\sigma _{0}}\right)^{3}\right]}\, ,
\label{electron density}
\end{align}
where $A$ is chosen in order to normalize the density: $A = 3N/(4 \pi \sigma ^{3}) [ \ln (1 + \exp ( r_c/\sigma _{0} )^{3} )]^{-1}$, $s$ is a displaced radial coordinate, $s (\bm{r},t ) =$ $ [ x^{2} + (y-d_y(t) )^{2} +  (z-d_z(t))^{2} ]^{1/2}$, and $\sigma_0$ is the equilibrium value of the electron spill-out effect.
In addition to the electron density, exact solutions for $V_{\rm H}(\bm{r},t)$ and $S(\bm{r},t)$ that satisfy respectively the Poisson equation and the continuity equation are given in  Ref.\ \citep{Hurst2014}. Here we report one of {those} solutions: $ S(\bm{r},t) =  \sigma / (2\dot{\sigma})s^2(\bm{r},t) + \dot{d}_{y}(z-d_y) + \dot{d}_{z}(z-d_z)$,
where the dot stands for the time derivative.
The associated electron mean velocity {is:}
\begin{align}
\bm{u} = \frac{\dot{\sigma}}{\sigma} x \bm{\widehat{x}} + \left[ \frac{\dot{\sigma}}{\sigma}\left( y - d_y \right) + \dot{d}_y \right] \bm{\widehat{y}}+\left[ \frac{\dot{\sigma}}{\sigma}\left( z - d_z \right) + \dot{d}_z \right] \bm{\widehat{z}}.
\label{velocity field}
\end{align}
Even {though} the average electron velocity diverges at infinity, it is physically acceptable because the relevant quantity is the electronic current, $\bm{j} = n \bm{u}$, which rapidly drops to zero outside the nanoparticle. Using the Eqs.\ \eqref{electron density} and \eqref{velocity field}, one can compute the total orbital magnetic moment as follows:
\begin{align}
\bm{M}(t) =\frac{1}{2} \int \bm{r} \times \bm{j}~ d\bm{r} = \frac{N}{2} \left[   \dot{d}_y d_z - \dot{d}_z d_y\right]\bm{\widehat{x}}.
\label{Total orbital magnetic moment}
\end{align}
Hence, one can in principle describe the generation of an orbital magnetic moment, provided that we excite both dipoles. Surprisingly, this expression is similar to the one derived in Ref.\ \cite{Battiato2014}: $\bm{M}(t) = -eN/2 \left[\bm{r}(t) \times \bm{\dot{r}}(t)\right]$, 
where the authors {computed} the classical magnetization in the framework of the classical Drude model. {Nonetheless, even though the results look similar, our work considers the fully self-consistent motion of an electron gas confined in a gold nanoparticle}.

Using the Ansatz \eqref{electron density} for the electron density, one can integrate the Lagrangian density \eqref{Lagrangian density gold nanoparticles}  over the whole space to obtain an analytical expression for the Lagrangian of the system:
\begin{align}
L &= \frac{-1}{N}\int \mathcal{L}_{D} (\bm{r},t)~d\bm{r} \nonumber  \\
& = \frac{M(a) \dot{\sigma}^{2}(t)}{2} - U(\sigma) + \frac{\dot{d}_y^{2}+ \dot{d}_z^{2}}{2} 
 -\frac{\Omega_{d}^{2}(\sigma)}{2}\left(d_y^2 + d_z^2 \right) \nonumber \\ 
 & \quad + K(\sigma) \left(d_y^2 + d_z^2 \right)^2 + d_y E_y + d_z E_z.
\label{The lagrangian}
\end{align}
The dipole terms are described by two coupled nonlinear oscillators whereas the breathing terms correspond to a fictitious particle of mass $M(a)$, where we introduce the small parameter $a =\exp\left(-r_{c}^{3}/ \sigma _{0}^{3} \right)$, moving in a time-dependent potential $U(\sigma)$.
In Eq.\ \eqref{The lagrangian}, the fictitious mass
\begin{equation}
M(a) = - \frac{\Gamma (5/3)\textrm{Li}_{5/3}(-1/a)}{\ln(1+1/a)}
\end{equation}
is given in terms of the gamma function $\Gamma (5/3) \simeq 0.90$ and the polylogarithm function $\textrm{Li}_{5/3}$ \footnote{The polylogarithm function is defined as ${\rm Li}_{p}(-1/a) = - [1/\Gamma(p)] \int_{0}^{\infty}dX\,X^{p-1}/(a\,e^{X}+1)$, where ${\rm Re}(p) > 0$, ${\rm Im}(a) = 0$, and $1/a > -1$.}. The multiplicative factor $-(1/N)$ was introduced in Eq.\ \eqref{The lagrangian} for convenience of notation. The other terms in Eq.\ \eqref{The lagrangian} are the pseudo-potential
\begin{align}
&U(\sigma)
 = \frac{f_{\rm B}(a)}{\sigma ^{2}} + \frac{N^{2/3}f_{\rm F}(a)}{\sigma ^{2}} - \frac{N^{1/3}f_{X}(a)}{\sigma } - \frac{\beta f_{X'}(a)}{N^{1/3} \sigma} \nonumber \\
& \quad +{}\!\! \frac{f_C(a) \sigma^2}{N} - \frac{f_{C'}(a) \sigma}{N} - \frac{f_{C''}(\sigma)}{N} + \frac{N f_{ee}(a)}{\sigma} - \frac{Nf_{ei}(\sigma)}{R}
\label{pseudo_pot}
\end{align}
and the functions
\begin{align}
\Omega _{d}^{2}(\sigma)
=& \frac{N}{R^{3} \ln(1+1/a)} \bigg{\{} \frac{R^{3}}{\sigma ^{3}} + \ln(1+a) \nonumber \\
& - \ln \left[1+a\exp(R^{3}/\sigma^{3}) \right]\bigg{\}},
\label{linear dipole freq}
\end{align}
\begin{align}
K(\sigma) = \frac{9NRa}{40 \ln(1+1/a)\sigma^{6}}~\frac{\exp(R^{3}/\sigma^{3})}{\left[1+a\exp(R^{3}/\sigma^{3})\right]^{2}},
\end{align}
which are both positive definite. 
Equation \eqref{The lagrangian} was obtained after some tedious algebra; {the details of this can be found in}
Ref.\ \cite{Hurst2014}. The only difference with the original derivation is that we have two dipole variables $d_y(t)$ and $d_z(t)$ in our Lagrangian instead of one. Mathematically, this difference appears via the substitution {of $(d_y^2+d_z^2 )^{1/2}$  for $d_z$}.  \\
\begin{figure}[t!]
\includegraphics[width=0.8\linewidth]{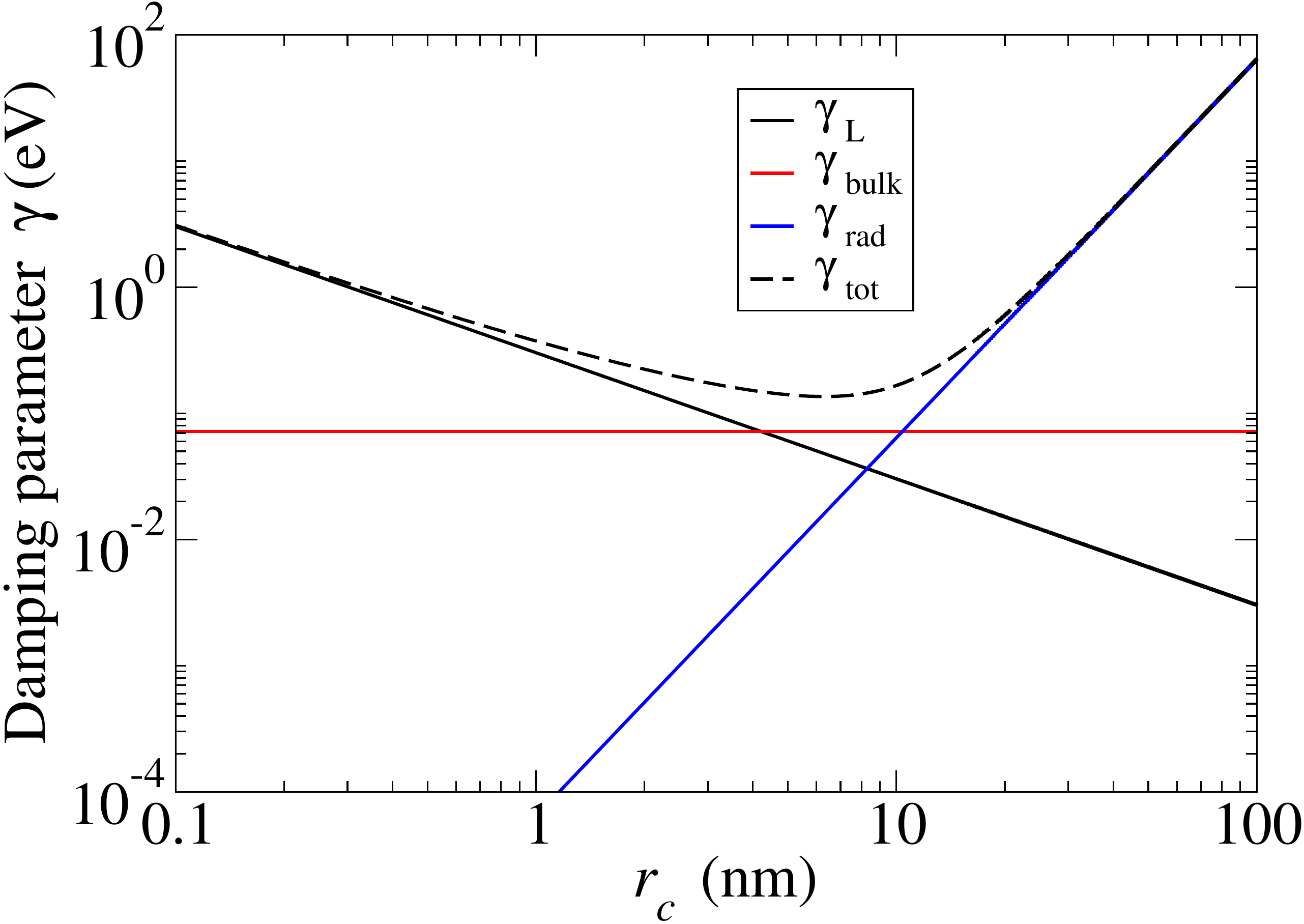}
\caption{(Color online)  Different damping parameters as a function of the nanoparticle's radius: Landau damping $\gamma_{\rm L}$ (black curve), nonradiative Ohmic losses $\gamma_{bulk}$ (red curve), radiative losses $\gamma_{rad}$ (blue curve) and the total damping $\gamma=\gamma_{\rm L} + \gamma_{bulk} + \gamma_{rad}$ (black dashed curve).}
\label{fig damping}
\end{figure}
The quantities $f_{\rm B}(a)$, $f_{\rm F}(a)$, $f_{X}(a)$, $f_{X'}(a)$, $f_{C}(a)$, $f_{C'}(a)$, $f_{C''}(\sigma)$, $f_{ee}(a)$ and $f_{ei}(\sigma)$, which appear in the pseudo-potential \eqref{pseudo_pot}, are given explicitly in Ref.\ \citep{Hurst2014} (supplementary material). They are related respectively to the Bohm potential, Fermi pressure, exchange energy (LDA), gradient correction to the exchange energy, electron-electron and electron-ion Hartree interaction terms. All these functions are positive, as well as the fictitious mass $M(a)$, in accordance with the role played by the Bohm, Fermi and electron-electron terms, which are repulsive, and by the exchange and the electron-ion terms, which are attractive. The correlation terms have both an attractive and a repulsive part.
The quantity $\Omega _{d}^{2}(\sigma)$ corresponds to the second order term in the development of the electron-ion interacting energy, whereas $K(\sigma)$ corresponds to the fourth order.\\
\begin{figure}[t]
\includegraphics[width=0.8\linewidth]{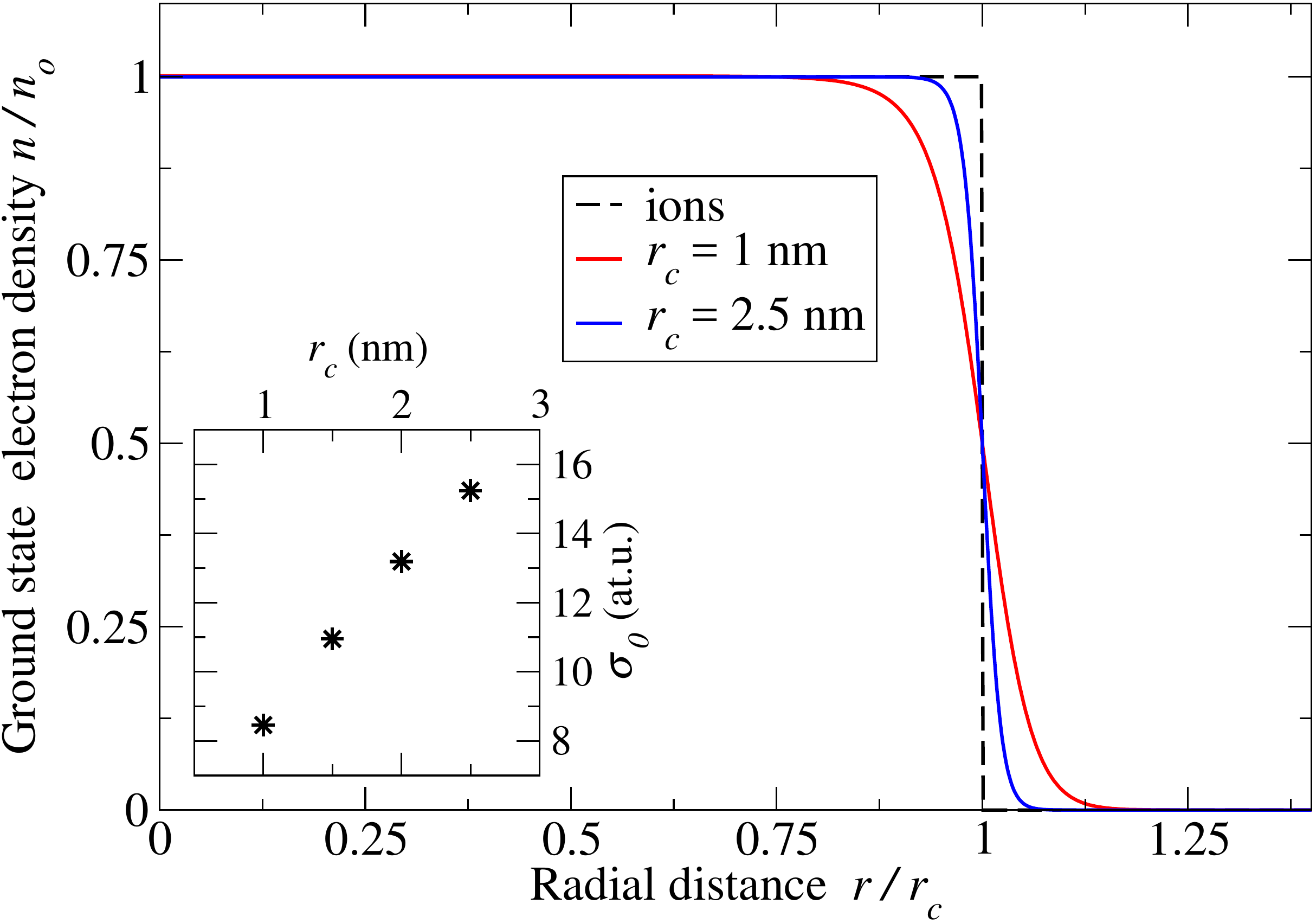}
\caption{(Color online) Radial profile of the ground state electronic density computed from Eq.\ \eqref{electron density}, for $r_c=1$ (red curve) and $r_c=2.5$ nm (blue curve). The dashed curve corresponds to the ion density. The inset {displays}
the values of $\sigma_0$ for different {radii} $r_c$ given in Table \ref{Tab-ground-state-value}.}
\label{fig ground state density}
\end{figure}
Using the Euler-Lagrange equation for $L$ we obtain the following equations of motion:
\begin{align}
\ddot{\sigma} &= \frac{1}{M(a)}\big[ - \frac{dU(\sigma)}{d\sigma} - \Omega_{d}(\sigma)\frac{d\Omega _{d}(\sigma)}{d\sigma} \left( d_y^2 + d_z^2\right)  \nonumber \\
& \quad  \quad +\frac{dK(\sigma)}{d\sigma} \left(d_y^2 + d_z^2 \right)^2\big],
\label{sigma equation} \\
\ddot{d}_y &=  - \Omega _{d}^{2}(\sigma)d_y + 4 K(\sigma) d_y\left(d_y^2 + d_z^2 \right)-\gamma \dot{d}_y + E_y,
\label{dy equation}\\
\ddot{d}_z &=  - \Omega _{d}^{2}(\sigma)d_z + 4 K(\sigma) d_z\left(d_y^2 + d_z^2 \right) - \gamma \dot{d}_z + E_z.
\label{dz equation}
\end{align}
Equation \eqref{sigma equation} describes the breathing dynamics of the electron cloud, whereas Eqs.\ \eqref{dy equation} and \eqref{dz equation} describe the dynamics of the center of mass of the electrons. The equations of motion of the dipoles consist of two nonlinearly coupled oscillators. In the linear regime both dipoles are decoupled and evolve as independent harmonic oscillators. This is in agreement with previous studies that predict a harmonic behavior in the linear regime for the center of mass of an electron gas confined in metallic nanoparticles \cite{Quinten2011,Weick2005}. In the nonlinear regime this property does not hold anymore and all the dynamical variables are coupled to each other.

We have also introduced a phenomenological damping term $\gamma = \gamma_{bulk} + \gamma_{rad} + \gamma_{L}$ in the dipole dynamics to simulate dissipative processes. It consists of three different components: (i) $\gamma_{bulk}$ describes the nonradiative Ohmic losses; {here} we take the bulk value for gold $\gamma_{bulk}=0.072$ eV \cite{Derkachova2016}, (ii) $\gamma_{rad} = 2 \Omega_d^4r_c^3/(3c^3)$ describes the radiative losses, which increase with the size of the nanoparticle  \cite{Brandstetter-Kunc2015}, {and} (iii) $\gamma_{L} \simeq 0.33 \, v_{\rm F}/r_c$ corresponds to the nonradiative Landau damping \cite{Li2013,Brandstetter-Kunc2015} that scales as the inverse of the nanoparticle radius.  In Fig. \ref{fig damping} we plot the different damping channels as a function of the nanoparticle radius. For small nanoparticles ($r_c < 5-10$ nm), the damping is mainly dominated by the Landau damping and the nonradiative Ohmic losses. On the contrary, for large nanoparticles ($r_c > 20$ nm), the damping is dominated by radiative losses that are large enough to considerably reduce the amplitude of the plasmon oscillations.
\begin{figure}[t]
\includegraphics[width=0.8\linewidth]{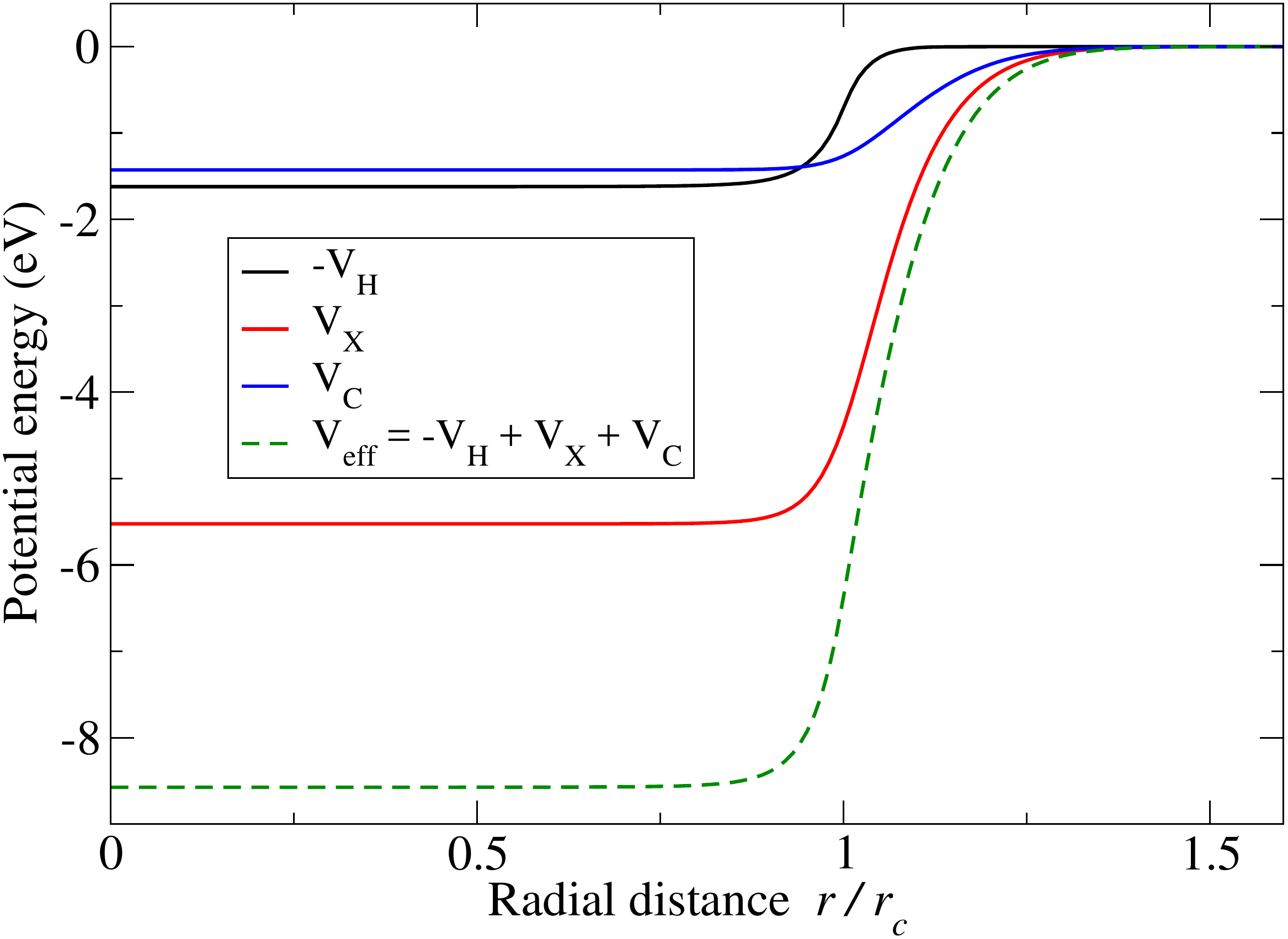}
\caption{ (Color online) Radial profile of the self-consistent ground state potentials: Hartree (black curve), exchange (red curve) and correlation (blue curve), for $r_c = 1$ nm. The dashed green curve corresponds to the effective potential $V_{eff} = -V_{\rm H} + V_{X} + V_{C}$ felt by the electrons.}
\label{fig ground state potential}
\end{figure}

\section{Linear regime and orbital magnetic moment generation}

The ground state density is obtained by setting
the center of mass variables to zero, i.e.\ $d_y=d_z=0$ and the bordering of the electron density to $\sigma = \sigma_0$. The latter value corresponds to the minimum of the pseudo potential $U\left( \sigma \right)$ in Eq.\ \eqref{pseudo_pot}. 
Several values of $\sigma_0$ are given in the Table \ref{Tab-ground-state-value} for different nanoparticle sizes. The corresponding ground state densities and potentials are plotted, respectively, in Figs.\ \ref{fig ground state density} and \ref{fig ground state potential}. We notice that the electron density spreads at the border of the nanoparticle as expected. Moreover, the bordering parameter $\sigma_0$ is proportional to the size of the nanoparticle and therefore the spreading of the electron density is almost the same for nanoparticles of different sizes. The exchange potential is three times larger than the Hartree and the correlation potentials, something that was already observed for thin metal films with DFT calculations \citep{Crouseilles2008}.
We limit our study to nanoparticles between $1$ nm and $2.5$ nm. The lower value is {constrained} by the fact that for smaller nanoparticles quantum effects play a significant role in the electron dynamics \cite{Weick2006}. For the upper limit, we are {constrained} by our Ansatz for the electron density that only includes the breathing and the dipole modes. Indeed multipolar  modes should be also considered for larger nanoparticles \cite{Kreibig1995}.

 Our model predicts the following plasmon resonance $\Omega_d(\sigma_0) = \omega_{\rm Mie} \sqrt{1 - \ln \left(2 \right) / \ln \left( 1 + \exp \left( r_c^3 / \sigma_0^3 \right) \right)}$,  which was  obtained by evaluating the linear dipole frequency \eqref{linear dipole freq} at $\sigma = \sigma_0$. Due to the spatial inhomogeneity of the electron density, the plasmon resonance increases with the size of the nanoparticle and in the limit of large nanoparticle, i.e. $r_c/\sigma_0 \gg 1$, tends to the bulk Mie frequency $\omega_{\rm Mie} = \omega_p / \sqrt{3} $ \citep{Kreibig1995}. Thus, our model reproduces the expected blue shift of the resonant dipole frequency \cite{Brechignac1993} that cannot be reproduced with a local Mie theory \cite{ruppin1973}.
\begin{table}[b]
\begin{ruledtabular}
\caption{ \label{Tab-ground-state-value}
Ground state and linear response parameters for gold nanoparticles of different sizes. {The different parameters given are: the radius of the nanoparticles $r_c$, the number of electrons $N$, the spreading of the electron density at the surface $\sigma_{0}$, the plasmon frequency $\Omega_d(\sigma_0)$,  and the damping constant $\gamma$.}}
\begin{tabular}{c c c c c}
$r_c$ [nm]  & $N$   & $\sigma_{0}$ [at.\,u.] & $\Omega_{d}(\sigma_0)$  [eV] &  $\gamma$  [eV] \\[0.1cm]
\hline
1 & 248 & 8.46 & 5.05 & 0.37\\
1.5 & 836 & 10.95 & 5.10 & 0.27 \\
2 & 1982 & 13.19 & 5.14  & 0.22\\
2.5 & 3870 & 15.24 & 5.15 & 0.19\\
\end{tabular}
\end{ruledtabular}
\end{table}

We use a continuous laser field to excite the electron dynamics. The laser field propagates in the $x$ direction and is described by the following electric field: $\bm{E}_L = E_0 \cos \left( \omega_L t \right) \bm{\widehat{y}} + E_0 \cos \left( \omega_L t -\phi \right) \bm{\widehat{z}}$. The phase parameter $\phi$ allows us to describe different light polarizations going from linear polarization ($\phi =0$) to circular left ($\phi =\pi/2$) or right ($\phi =-\pi/2$) polarization. We neglect the {spatial} variations of the electric field because the corresponding wavelength is much larger than the size of the nanoparticles. 

\begin{figure}[b]
\includegraphics[width=0.8\linewidth]{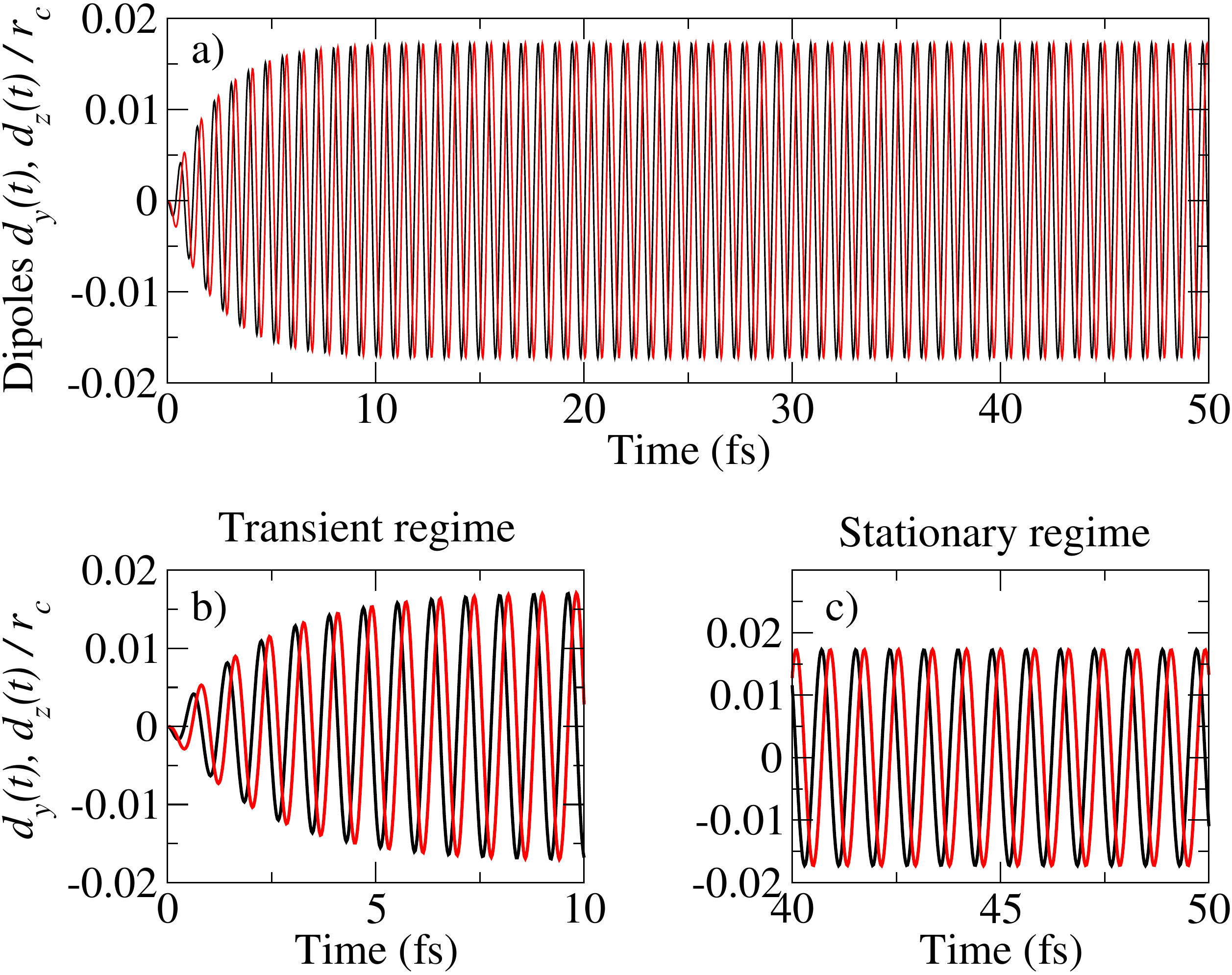}
\caption{(Color online) (a) Plot of the {calculated} dipole dynamics $d_y(t)$ (black curve) and $d_z(t)$ (red curve) for a gold nanoparticle with a radius $r_c=1$ nm. The system is excited with a circular right polarized laser field ($\phi=-\pi/2$) with an amplitude $E_0=6.2 \times 10^{8}$ V/m. Plots (b) and (c) are zooms taken at different times.}
\label{fig dipole_circular_light}
\end{figure}

\begin{figure}[t]
\includegraphics[width=0.8\linewidth]{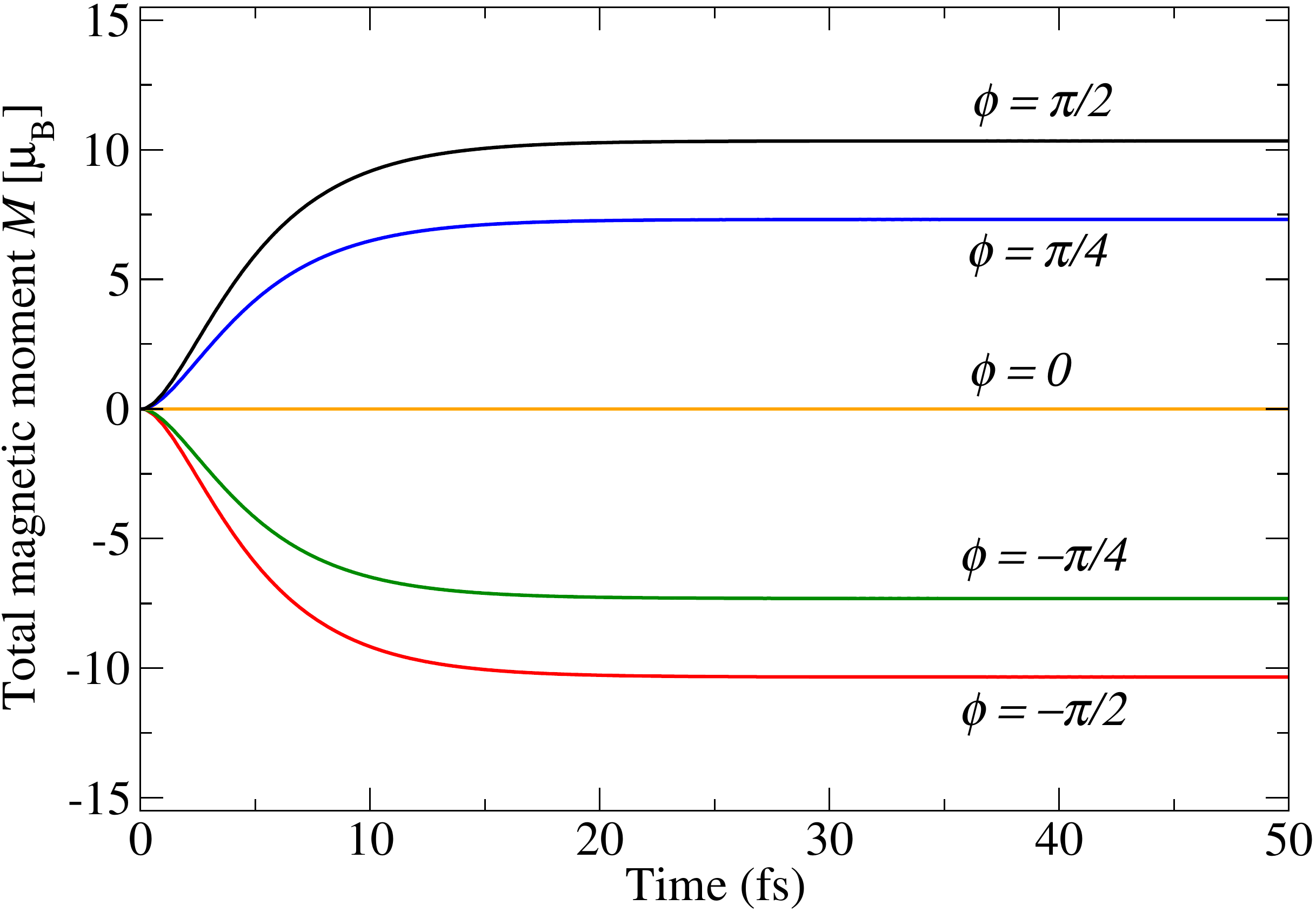}
\caption{(Color online) Time evolution of the total {laser-induced} orbital magnetic moment {in a Au nanoparticle} computed with formula \eqref{Total orbital magnetic moment} for different laser polarizations {$\phi$}. The simulation parameters {used} are the same as {those of} Fig.\ \ref{fig dipole_circular_light}.}
\label{fig orbital_magnetization}
\end{figure}

In a first simulation, we excite the system with a circular right polarized {field} of intensity $I_L = 5.1 \times 10^{10}$ W/cm$^2$, which corresponds to an electric field $E_0=6.2 \times 10^{8}$ V/m. In all the simulations we took the laser frequency equal to the resonant dipole frequency of the system, $\omega_L=\Omega_d(\sigma_0)$, given in Table  \ref{Tab-ground-state-value}. We have checked that with such an intensity we are in the linear regime. The dipole responses $d_y(t)$ and $d_z(t)$ are given in Fig.\ \ref{fig dipole_circular_light}. The results {were} obtained by solving the equations of motion \eqref{sigma equation}-\eqref{dz equation} with a Runge-Kutta method of order $4$. The dynamics is characterized by two regimes. During the first ten  femtoseconds, one observes a \textit{transient} regime in which the dipoles increase in amplitude. The typical timescale for this regime is given by the inverse of the damping parameter $\gamma$.  After that, the system reaches a stationary regime, where the dipoles are oscillating with a phase shift of $\pi/2$. We point out that, when one switches off the laser field, the dipoles behave as anharmonic damped oscillators, see Eqs.\ \eqref{dy equation}-\eqref{dz equation}. Therefore, they will be exponentially damped on a time scale given by the inverse of the damping parameter $\gamma$.\\
{Employing} Eq.\ \eqref{Total orbital magnetic moment}, we can directly compute the total orbital magnetic moment along the $x$ direction from the dipole responses $d_y(t)$ and $d_z(t)$. The results are given in Fig.\ \ref{fig orbital_magnetization} for different polarizations of the incoming laser field, $\phi = \pm \pi/2, ~\pm \pi/4$, and 0. We notice that in the case of a linearly polarized electric field, the total orbital magnetic moment remains zero. In {this} case one can check that both dipoles are oscillating in phase. However, if we use a circular right (black curve) or left (red curve) polarized {field}, then a net orbital magnetic moment is created in the system. {This magnetic moment}
increases during the transient regime until {it reaches} a stable value in the stationary regime. This situation corresponds to the case where the electric dipoles are oscillating {with a} phase {offset}
as pictured in Fig.\ \ref{fig dipole_circular_light}. Moreover, one observes an opposite effect for circular left and circular right polarizations. Finally, if we excite the system with an elliptically polarized electric field (here $\phi = \pm \pi/4$), then we still obtain a nonzero magnetic moment but smaller than the one obtained with {full} circular polarization. To summarize, in order to create a nonzero magnetic moment in gold nanoparticles, one has to excite the system at the resonant frequency of the surface plasmon. If one excites the system far from its resonance, then the dipoles will be significantly reduced as well as the magnetic moment. {This is simply due to the fact that the dipole dynamics are described by nonlinear damped oscillators.}\\
These observations are typical signatures of the inverse Faraday effect, where a part of the spin angular momentum of the light has been transferred to the electrons. An inverse Faraday effect has been predicted in {a} large gold nanoparticle \cite{Nadarajah2017} ($r_c=100$ nm) using a full-wave electrodynamics solver (Lumerical) combined with individual electron motions. {There,} the authors have shown that the inverse Faraday effect emerges from an ensemble of solenoid-like motions for each electron inside the nanoparticle. In the next section, we propose a different explanation of the origin of the inverse Faraday effect observed in our system.

\section{Mechanism of orbital magnetic moment generation}

According to Eq.\ \eqref{Total orbital magnetic moment}, the orbital magnetic moment is defined in terms of the electronic current density $\bm{j}$. The latter can be expressed as the product of the electron density and the electron mean velocity,
\begin{align}
\bm{j} = n\bm{u} = &n\left( \bm{r},t\right)\frac{\dot{\sigma}}{\sigma} x \bm{\widehat{x}} + n\left( \bm{r},t\right)\left[ \frac{\dot{\sigma}}{\sigma}\left( y - d_y \right) + \dot{d}_y \right] \bm{\widehat{y}} \nonumber \\
& +n\left( \bm{r},t\right)\left[ \frac{\dot{\sigma}}{\sigma}\left( z - d_z \right) + \dot{d}_z \right] \bm{\widehat{z}},
\end{align}
where the time-dependent electron density $ n\left( \bm{r},t\right)$ is defined in Eq.\ \eqref{electron density}. The time dependence of the electronic current density is embedded in the dipole and breathing variables as well as in their time derivatives.
It is straightforward to see that in the ground state the electronic current is zero. However, during the dynamics the spatial profile looks rather complicated especially near the surface of the nanoparticle.  

\begin{figure}[t]
\includegraphics[width=1.1\linewidth]{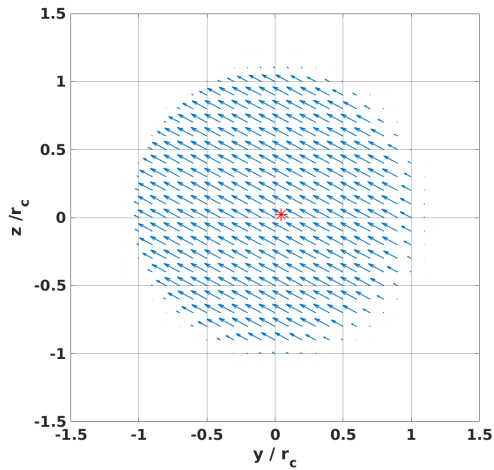}
\caption{(Color online) Plot of the current density vector field $j_{y}$ and $j_{z}$ in the plane $\left\{x=0\right\}$ at a given time. The red star near the origin represents the position of the center of mass of the electrons. The system is a gold nanoparticle with a radius $r_c=1$ nm and the {laser} excitation is circularly polarized with an amplitude $E_0=6.2 \times 10^{8}$ V/m.}
\label{fig electronic_current2D_at_time_t}
\end{figure}

In Fig.\ \ref{fig electronic_current2D_at_time_t} we plot the current density vector field at a given time using the values of the dipoles obtained in Fig.\ \ref{fig dipole_circular_light}. We only plot the $y$ and $z$ components of $\bm{j}$ in the plane defined by $\left\{x=0\right\}$ since the $x$ component is exactly zero in this plane. We notice that almost all the vectors point in the same direction. This direction is defined by the instantaneous laser field and is changing in time since we excite the system with a circularly polarized electric field. The current density is strongest at the center of the nanoparticle and decrease{s} rapidly at the border of the nanoparticle.
From this plot, it is not easy to understand the origin of the orbital magnetic moment. It would seems that if we sum up all the contributions  $\bm{j} \times \bm{r}$ in the integral of Eq.\ \eqref{Total orbital magnetic moment} one obtains zero, but this is not the case. Notably in Fig.\ \ref{fig electronic_current2D_at_time_t} the current density vector field is not centered around {the origin} but around the center of mass  of the electrons $\left( d_y(t),\, d_z(t)\right)$. The latter enscribes a small circle around the center of the nanoparticle during one pulse oscillation. Mathematically speaking, this explains why the integral in Eq.\ \eqref{Total orbital magnetic moment} is not zero but {adopts} a finite value that depends on both dipole variables, {$\bm{j}$ and $\bm{r}$.}

To have a better understanding of the underlying mechanism that is responsible {for} the generation of an orbital magnetic moment, let us define a time-averaged electron current density: $\langle \bm{j} \rangle = 1/T_{d} \int_t^{t+T_d} \bm{j}(t')~ dt'$. The time integration has to be done in the stationary regime over a full dipole period $T_{d}$.
\begin{figure}[t]
\includegraphics[width=0.8\linewidth]{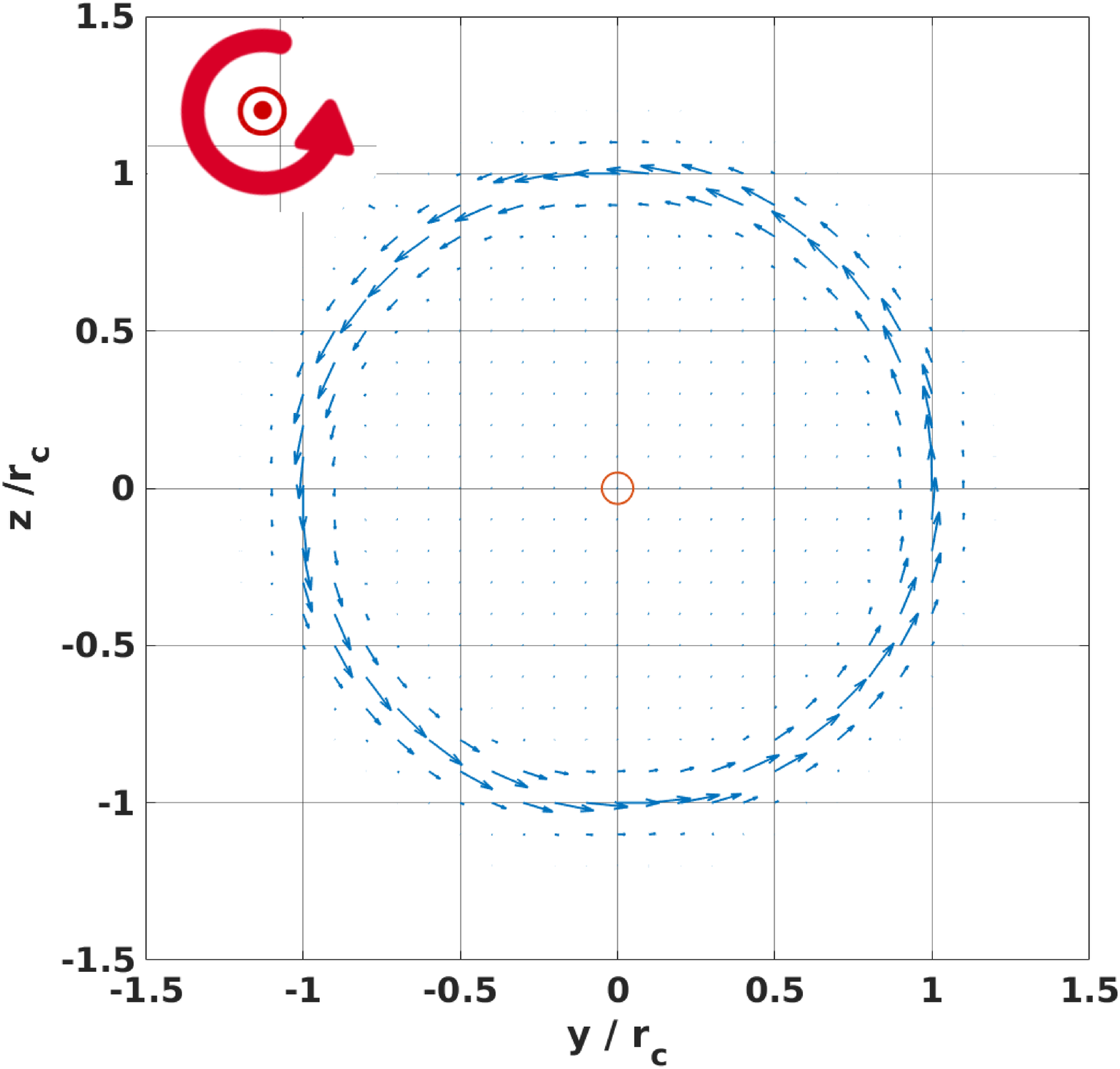}
\caption{(Color online) {Plot of the time-averaged} current density vector field $j_{y}$ and $j_{z}$ represented in the plane $\left\{x=0\right\}$. {The time average is} over one dipole period. The red line represents the successive positions of the center of mass of the electrons during one laser oscillation. {The simulated system is the same} as in Fig.\ \ref{fig electronic_current2D_at_time_t}.  The laser field propagates in the $x$ direction and is circularly left polarized as indicated in the top left-hand corner.}
\label{fig electronic_current2D_time_average}
\end{figure}
In Fig.\ \ref{fig electronic_current2D_time_average} we plot the time-averaged current density corresponding to the same simulation as {shown} in Fig.\ \ref{fig electronic_current2D_at_time_t}. The averaged current density vanishes everywhere except at the surface of the nanoparticle. Moreover, the current density is rotating around the $x$ axis. This structure emerges from the superposition of many current densities that are all pointing in different directions defined by the instantaneous laser field. On average, they cancel everywhere except at the surface of the nanoparticle,  because, as was mentioned before, each current is centered {around the oscillating} center of mass of the electrons. {Thus,} even though there are no real rotating surface currents, the system behaves as {if} that was the case. In the rest of this work, we will use the time average current density to evaluate the magnetic properties of the gold nanoparticles.

\begin{figure}[b]
\includegraphics[width=0.9\linewidth]{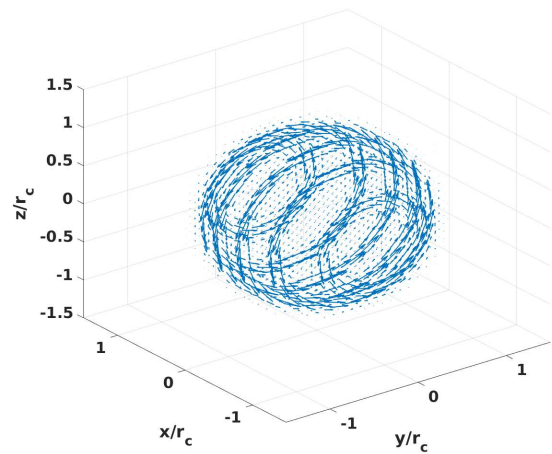}
\caption{(Color online) {Three-dimensional} plot of the {calculated} time-averaged current density vector field,  shown in the planes defined by: $x = \left\{0~ ;~ \pm 0.5r_c ~;~ \pm 0.8r_c \right\}$. The {simulated} system is the same as {that of} Figs.\ \ref{fig electronic_current2D_at_time_t} and \ref{fig electronic_current2D_time_average}. }
\label{fig 3d_current}
\end{figure}
\begin{figure}
\includegraphics[width=0.9\linewidth]{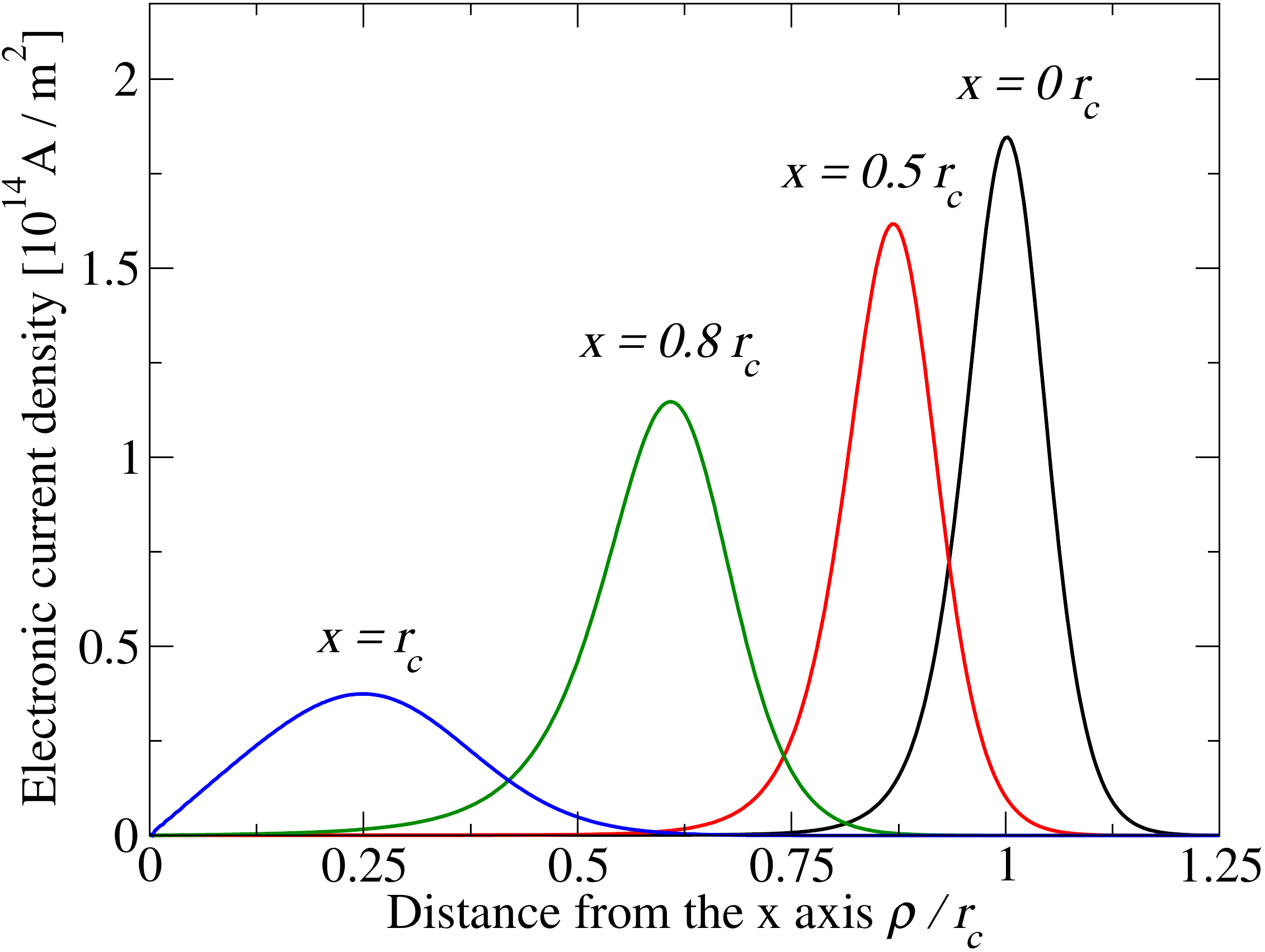}
\caption{(Color online) Radial profile as a function of the distance from the $x$ axis [$\rho = ({y^2 + z^2})^{1/2}$] of the current density {shown} in Fig.\ \ref{fig 3d_current}, for different $y-z$ planes defined by $x=0$ (black curve), $x=0.5\,r_c$ (red curve), $x=0.8\,r_c$ (green curve), and $x=r_c$ (blue curve).}
\label{fig Electronic_current_vs_r}
\end{figure}

In Fig.\ \ref{fig electronic_current2D_time_average} we have shown that the instantaneous current density can be mapped {onto} rotating surface currents. But that was only done in a given plane, defined by $x=0$. In Fig.\ \ref{fig 3d_current}, we represent the {time-averaged} current density over the {whole} nanoparticle. We {can recognize}  that the above assertion remains valid {for the whole} nanoparticle. This picture is particularly relevant to understand the behavior of the orbital magnetic moment shown in Fig.\ \ref{fig orbital_magnetization}. If we change the polarization of the light from circular-right to circular-left, then the current will simply flow in the opposite direction and the induced magnetic moment will change its sign.

Finally, in Fig.\ \ref{fig Electronic_current_vs_r} we analyze the intensity of the current density versus the radial distance from the $x$ axis [$\rho = ( y^2 + z^2 )^{1/2}=( r^2-x^2 )^{1/2}$] in four different $y-z$ planes defined by $x=\left\{0\right\}$, $\left\{0.5\,r_c \right\}$, $\left\{0.8\,r_c\right\}$, and
$\left\{r_c\right\}$, respectively. The radial profile is obtained by averaging the current density over the cylindrical angle. The current density is peaked around the surface of the nanoparticle, as expected. For instance, for $x=0.5\,r_c$, the peak is observed at $\rho_{max}=( r_c^2-x^2 )^{1/2} \approx 0.87 \, r_c$.
We {further note} that the current density is maximal in the plane $x=0$ and decreases progressively when one moves along the $x$ axis. The value of the electric current density is of the order of $10^{14}$ A/m$^2$, which seems to be reasonable because it corresponds approximatively to a single electron crossing a surface of $1$ nm$^2$ each femtosecond. However, this value depends mainly on the size of the nanoparticle and on the intensity of the laser excitation. This issue will be discussed in the next section.


\section{Nonlinear regime and size dependence}
\begin{table}[b]
\begin{ruledtabular}
\caption{
\label{Tab-quantities-vs-r}
{Given are the} maximal current density $j_{max}$, total magnetic moment $M$, and magnetic field $B$ calculated for different nanoparticle sizes $r_c$. The {applied} laser field has a circular right polarization and an intensity of $5.1 \times 10^{10}$ W/cm$^2$. }
\begin{tabular}{c c  c  c c}
$r_c$ [nm] & $j_{max}$ [$10^{14}$A/m$^2]$  & $M$ $\left[\mu_{\textrm{B}} \right]$  & $B$ $(r\!=\!0) $ [T]  \\[0.1cm]
\hline
1 & 1.85 & 10.4 & 0.019 \\
1.5 & 3.41 & 65.2 & 0.030  \\
2 & 4.93 & 228.1 & 0.053  \\
2.5 & 6.45 & 599.9 & 0.071 \\
\end{tabular}
\end{ruledtabular}
\end{table}

All the results discussed in the previous sections were obtained for gold nanoparticles with a radius of $1$~nm and for laser excitations with an intensity of $5.1 \times 10^{10}$ W/cm$^2$. In this section, we study the influence of the nanoparticle size and {of} the laser intensities on the magnetic properties of the gold nanoparticles. 

In Table  \ref{Tab-quantities-vs-r}, $j_{max}$ denotes the maximal value of the time-averaged current density, obtained at the surface of the nanoparticle for {plane} $x=0$. We also provide the total magnetic moment $M$ and the magnetic field $B$ at the center of the nanoparticles for four different sizes. 
The magnetic {field} is calculated with the Biot-Savart law,
\begin{align}
\bm{B}(\bm{r},t)  = \frac{\mu_0}{4 \pi} \int \frac{ \bm{j} (\bm{r}',t')  \times \left( \bm{r} - \bm{r}'\right)}{|\bm{r} - \bm{r}'|^3} d\bm{r}'.
\label{biot savart field}
\end{align}
The numerical integration of Eq.\ \eqref{biot savart field} {gives} a static magnetic field $B$ at the center of the nanoparticle, such that one can use the time-average current density $\langle \bm{j}(\bm{r}) \rangle$ instead of the current density $ \bm{j}(\bm{r},t) $.
We give the value of the magnetic field only at the center of the nanoparticle because it {reaches} it's largest value at that particular position. The laser intensity remains equal to $5.1 \times 10^{10}$ W/cm$^2$ so that we are in the linear regime. 
The total magnetic moment and the magnetic field at the center of the nanoparticle are along the $x$ direction and can be positive or negative depending {on whether} the surface currents are {rotating} clockwise or counter-clockwise.

\begin{figure}[t]
\includegraphics[width=0.95\linewidth]{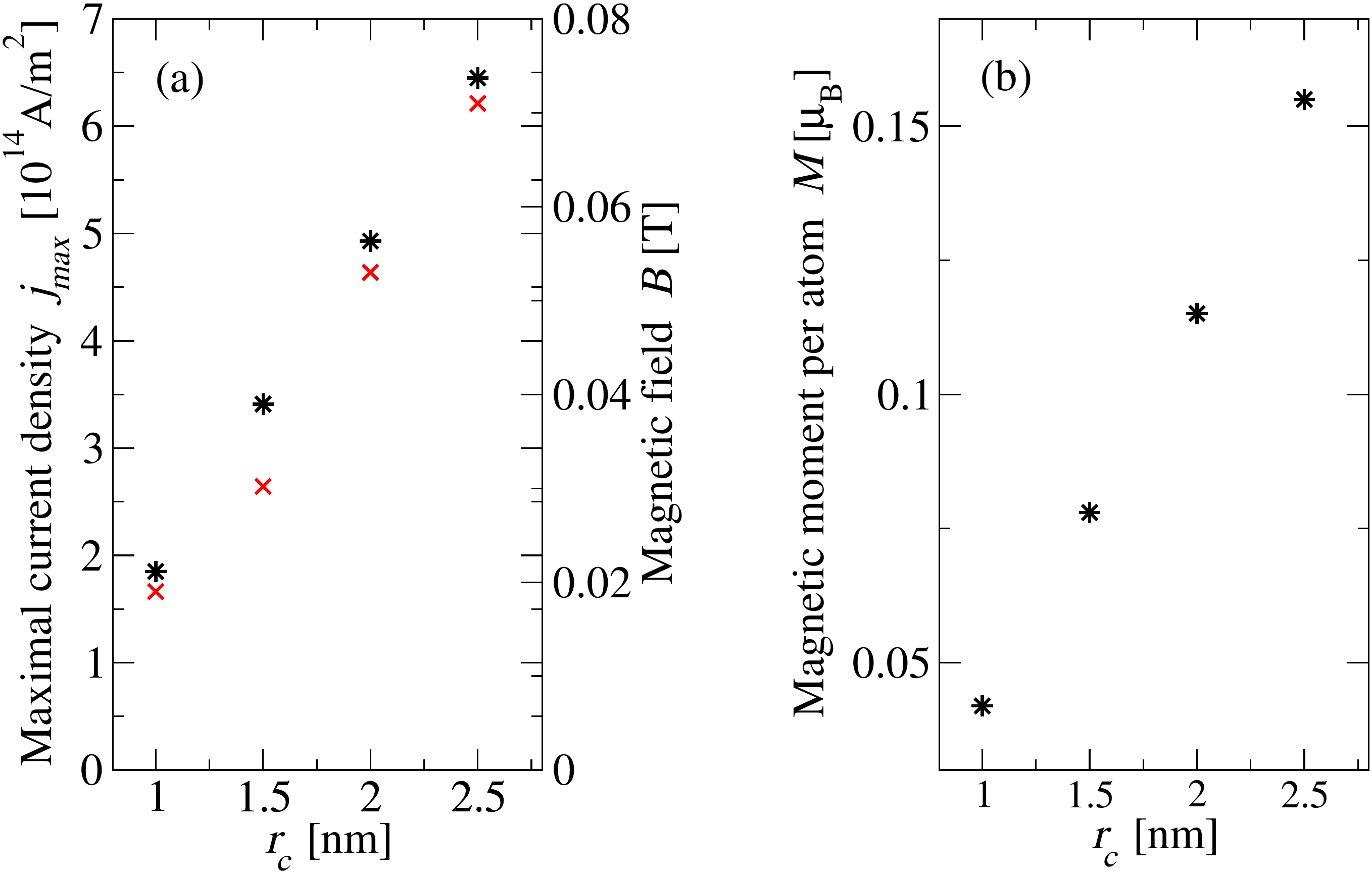}
\caption{(Color online).  (a): {Calculated} maximal current density (black stars) and magnetic field (red crosses) as a function of the nanoparticle radius. (b): Computed magnetic moment per atom as a function of the nanoparticle radius. 
}
\label{fig_quantity_vs_r}
\end{figure}

In Fig.\ \ref{fig_quantity_vs_r} we investigate the size dependence of the different quantities given in Table \ref{Tab-quantities-vs-r}. 
We {observe} that both the maximal current density and the magnetic moment per atom increase linearly with the radius of the nanoparticle. A similar trend is observed for the magnetic field at the center of the nanoparticle. The behaviors of the magnetic field and the magnetic moment can be understood from the behavior of the maximal current density by considering the integral over the surface of the nanoparticle in Eqs.\ \eqref{Total orbital magnetic moment} and \eqref{biot savart field}. Note that the total magnetic moment scales as $r_c^4$ because the number of atoms in the nanoparticle scales as $r_c^3$. This explains the {large  increase} of the total magnetic moment {seen} in Table 
\ref{Tab-quantities-vs-r}.  Our model predicts an increase of the magnetic moment and the magnetic field with an increase of the size of the nanoparticle. However, this will not happen indefinitely. The reason, which was already mentioned before, is that for larger nanoparticles ($r_c > 30-40$ nm) we have a strong damping due to radiative losses, see Fig.\ \ref{fig damping}. The latter may considerably reduce the amplitude of the dipoles and hence the amplitude of the magnetic moment and the magnetic field.

\begin{figure}[t]
\includegraphics[scale=0.3]{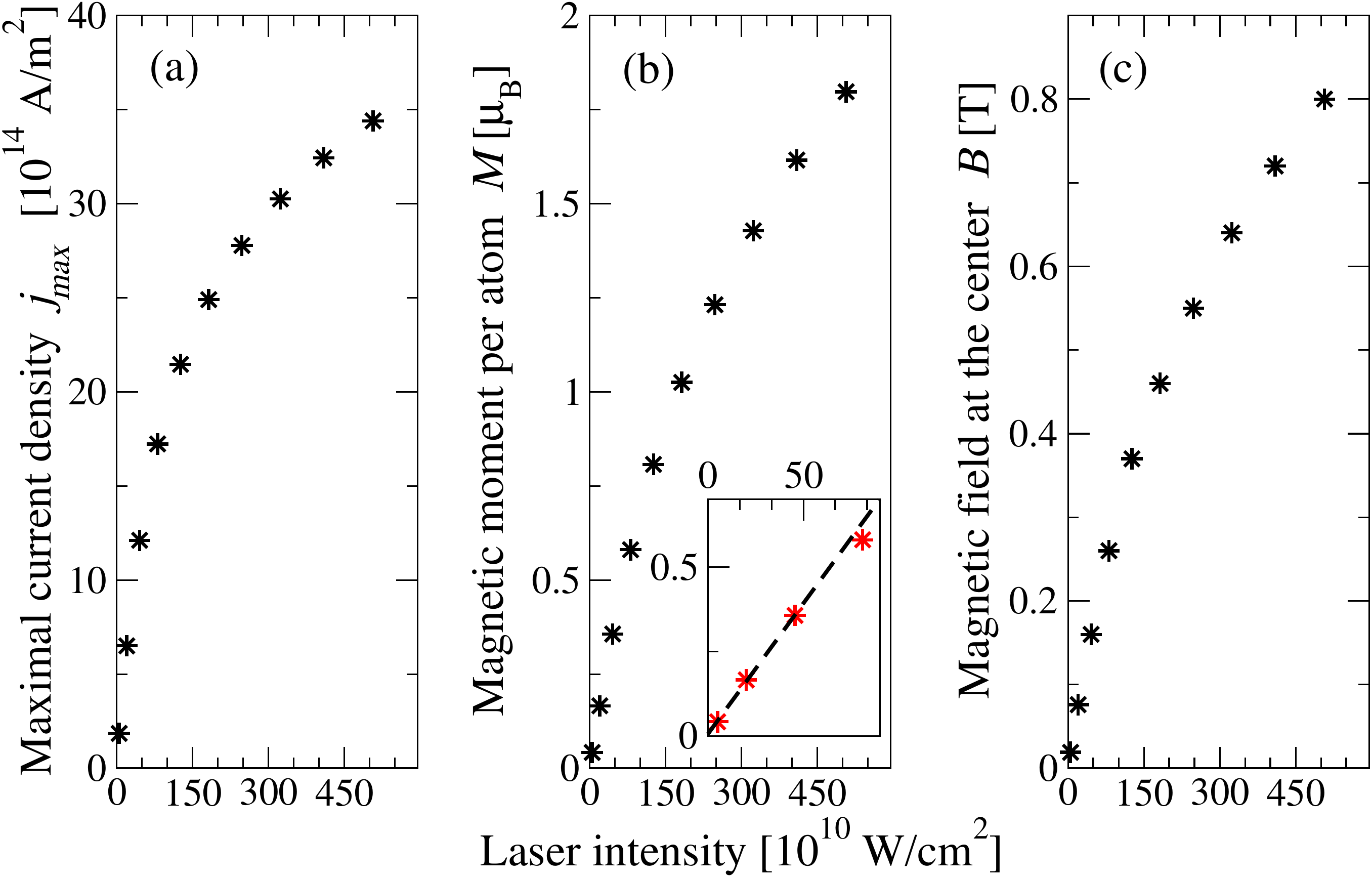}
\caption{(Color online). (a) Maximal current density, (b), magnetic moment per atom, and (c) the magnetic field at the center of the nanoparticle, {computed} as a function of the laser intensity. The inset plot {in (b)} is a linear fit of the smallest values of the magnetic moment per atom. The system is a gold nanoparticle with a radius $r_c=1$ nm.}
\label{fig_quantity_vs_power}
\end{figure}

In Fig.\ \ref{fig_quantity_vs_power} we study the influence of the laser intensity on the maximal current density, the magnetic moment per atom, and the magnetic field, {respectively.} The system is in the linear regime for the smallest intensity ($5.1 \times 10^{10}$ W/cm$^2$), but not for the highest laser intensity ($5.07 \times 10^{12}$ W/cm$^2$).
One observes that all quantities increase with the laser field, {simply because 
the electrons absorb more energy from the laser, which increases the dipole motion and thus the maximal current density}. The total magnetic moment and the magnetic field follow the same trend as the current density. For small intensities, the magnetic moment increases linearly with the laser intensity and hence with $E_0^2$, see Fig.\ \ref{fig_quantity_vs_power}(b). This again demonstrates that the generation of an orbital magnetic moment in our system is an {orbital} inverse Faraday effect.
{Note} that the magnetic field created at the center of the nanoparticle is static during the {duration of the} laser pulse and can reach considerable values ($0.1-0.8$ T). In the nonlinear regime, the increase of the induced magnetic quantities starts leveling off. 
Simulations performed at higher laser intensities (not shown here) reveal a saturation of all such quantities to a maximal value.\\
The quantities shown in Fig.\ \ref{fig_quantity_vs_power} are probably overestimated for the largest laser intensity, especially {when} higher-order multipolar plasmon modes (quadrupole, octupole) start to play a significant role in the electron dynamics.
The reason is that our model is based on the assumption that the electron density remains isotropic during all the dynamics. {Although} this assumption can be justified in the linear regime, it is not necessarily valid in the strongly nonlinear regime. We expect that a more general description of the electron density, taking into account higher-order {multipolar} modes, will be more accurate in the nonlinear regime. Indeed the latter will break the spherical symmetry of the electron density and thus probably  reduce the rotating surface currents.  Moreover, other nonlinear effects such as ionization or generation of solitons \cite{Koshelev2017}, which cannot be described with our Ansatz, may lead to a different electron dynamics in the nonlinear regime.
\\

\section{Discussions and Conclusions}

{We have} used a QHD model to show that, under the action of a circularly polarized laser field, gold nanoparticles can build up a static magnetic moment. {We have} shown that the induced magnetization per atom is proportional to the radius of the nanoparticle. The corresponding physical mechanism can be understood by analyzing the time-averaged electron current density, which exhibits rotating surface currents.
The latter arise from collective effects described by the dipole and the breathing dynamics of the electron cloud. We have shown that this mechanism exhibits all the properties of a classical inverse Faraday effect: (i) the induced magnetization is static and reverses its sign when changing the light polarization from circular right to circular left, (ii) there is no induced magnetic moment {for} a linear polarization, {and} (iii) for small laser intensities the induced magnetization scales as the square of the electric field.
We emphasize here the importance of the finite size of the system, which constitutes an essential ingredient for the generation of the magnetic moment. We would like to stress that in order to create a magnetic moment in gold nanoparticles, one has to excite the system at the resonant frequency of the surface plasmon. {For excitations far off the plasmon resonance the magnetic moment remains close to zero.}

In our model, a magnetic moment is created in gold nanoparticles by the collective motion of the electrons that interact with a circularly polarized laser field. {This is in contrast with an earlier approach \cite{Nadarajah2017}, where the authors analyzed that the magnetic moment emerges from an ensemble of independent solenoid-like motions for each electron. The QHD model employed in the present work goes beyond the independent and free-electron approximation by taking into account the main quantum many-body effects, such as the Hartree potential and the exchange and correlation effects. Using the same laser intensity, we predict an electronic current density four orders of magnitudes larger than in Ref.\ \cite{Nadarajah2017}. Although this difference is important, it is most probably due the different approach used in the two models. Indeed, in Ref.\ \cite{Nadarajah2017}, the authors have assumed that the charge distribution in the nanoparticle stays uniform during the entire laser pulse and hence they neglect any redistribution of the electric field due to other nonlinear phenomena. On the contrary, in our model, the electric current responsible for the creation of an orbital magnetization is caused by the combination of the dipole motions (surface plasmons) and the inhomogeneity of the electron density at the surface of the nanoparticle (spill-out). 

Summarizing, we have shown that surface plasmons support the generation of an orbital angular momentum in gold nanoparticles. This  phenomenon corresponds to a transfer of the spin angular momentum of the light to the electronic orbital degree of freedom in the nanoparticle through the plasmonic orbital inverse Faraday effect. 
As a result, a static magnetic field is created inside the nanoparticle during the laser pulse.  In future studies, it would be interesting to study other geometries such has nano-rings, since a resonant inverse Faraday effect was recently predicted in such nanostructures \cite{Koshelev2015,Koshelev2017}. The computed induced magnetic moments in the nanoparticle are quite large, of about 0.35 $\mu_{\rm B}$/atom for a laser intensity of $45 \times 10^{10}$ W/cm$^2$. Our study focused on gold nanoparticles but in principle other materials, such as silver or aluminium, could be investigated as well within the present theory. The decisive point for the magnetic moment generation is that the material supports a strong plasmonic response at the driving laser frequency. 
The thus-generated  magnetic field could be employed as a new approach to achieve ultrafast plasmon-assisted all-optical switching in suitable systems such as core/shell nanoparticles, supported gold discs or gold rings with a magnetic core inside. 

\acknowledgments{
We gratefully acknowledge valuable discussions  with F.\ Haas, M.\ Berritta, A.\ Dmitriev, and V.\ Kapaklis. 
This work has been financially supported by the European Union's Horizon2020 Research and Innovation Programme under Grant agreement No.\ 737709 (FEMTOTERABYTE, 
http://www.physics.gu.se/femtoterabyte), the Knut and Alice Wallenberg Foundation (Contract No.\ 2015.0060), the Swedish Research Council (VR), and the Swedish National Infrastructure for Computing (SNIC).
}

\bibliographystyle{apsrev4-1}

\bibliography{biblio}{}

\end{document}